\documentclass[12pt,a4paper,final]{iopart}

\usepackage{iopams}  
\usepackage{graphicx}
\usepackage[breaklinks=true,colorlinks=true,linkcolor=blue,urlcolor=blue,citecolor=blue]{hyperref}

\usepackage{amsfonts}

\expandafter\let\csname equation*\endcsname\relax
\expandafter\let\csname endequation*\endcsname\relax
\usepackage{amsmath,bm,amssymb}

\usepackage{braket}

\usepackage[numbers,sort&compress]{natbib}

\begin{document}
	
	\title[Magnetic force theory combined with quasi-particle self-consistent GW method ]{Magnetic force theory combined with quasi-particle self-consistent GW method}

	\author{Hongkee Yoon$^{1}$, Seung Woo Jang$^{1}$, Jae-Hoon Sim$^{1}$, Takao Kotani$^{2}$ and Myung Joon Han$^{*1},^{3}$}

	\address{$^{1}$Department of Physics, KAIST, 291 Daehak-ro, Yuseong-gu, Daejeon 34141, Republic of Korea}
	\address{$^{2}$Department of Applied Mathematics and Physics, Tottori University, Tottori 680-8552, Japan}
	\address{$^{3}$KAIST Institute for the NanoCentury, Korea Advanced Institute of Science and Technology, Daejeon 34141, Korea}
	\ead{mj.han@kaist.ac.kr}

	\begin{abstract}
We report a successful combination of magnetic force linear response theory with quasiparticle self-consistent GW method. The self-consistently determined wavefunctions and eigenvalues can just be used for the conventional magnetic force calculations. While its formulation is straightforward, this combination provides a way to investigate the effect of GW self-energy on the magnetic interactions which can hardly be quantified due to the limitation of current GW methodology in calculating the total energy difference in between different magnetic phases. In ferromagnetic $3d$ elements, GW self-energy slightly reduces the $d$ bandwidth and enhances the interactions while the same long-range feature is maintained. In antiferromagnetic transition-metal monoxides, QSGW significantly reduces the interaction strengths by enlarging the gap. Orbital-dependent magnetic force calculations show that the coupling between $e_g$ and the nominally-empty $4s$ orbital is noticeably large in MnO which is reminiscent of the discussion for cuprates regarding the role of Cu-$4s$ state. This combination of magnetic force theory with quasiparticle self-consistent GW can be a useful tool to study various magnetic materials.

	\end{abstract}
	
	\vspace{2pc}
	\submitto{\JPCM}

\section{Introduction}

Understanding various magnetic phases of matter has long been a central issue in physics. On the one hand, controlling and utilizing magnetic orders are the key for memory device application \cite{gallagher_development_2006,parkin_magnetic_2008,fert_skyrmions_2013}. On the other, competition and cooperation of magnetic degree of freedom with other ingredients of solids can stabilize new quantum states of matter such as high-T$_C$ superconductivity and quantum spin liquid \cite{scalapino_common_2012,lee_doping_2006,fradkin_colloquium_2015,balents_spin_2010,zhou_quantum_2017}. There has been a significant advance in both experiment and theory toward measuring and characterizing the magnetic order and interaction. Within the first-principles theoretical framework, there are basically two different ways widely used. One is to compare the calculated total energies of both ground state and meta-stable spin-ordered states \cite{oguchi_band_1983,oguchi_transitionmetal_1984}. By mapping this energy difference onto a well-defined spin model such as Heisenberg spin Hamiltonian, one can determine the magnetic couplings.

Another way is to resort to so-called magnetic force linear response theory (MFT) \cite{oguchi_magnetism_1983,liechtenstein_local_1987}. One of the advantages of using MFT is that, without any supercell calculation, it estimates both short- and long-range interactions. Also, as a conceptually different approach, it can provide complementary information. In fact, the idea of MFT is well consistent with the principle of inelastic neutron scattering, the most standard experimental technique to measure the magnetic interactions. Further, MFT can be utilized in predicting the magnetic ground state especially when the system carries a well localized magnetic moment \cite{yoon_reliability_2018}. Due to these advantages, MFT has been developed persistently \cite{oguchi_magnetism_1983,liechtenstein_local_1987,antropov_spin_1996,liechtenstein_exchange_1984,antropov_exchange_1997,han_electronic_2004,yoon_reliability_2018,solovyev_effective_1998-1,katsnelson_first-principles_2000,bruno_exchange_2003,udvardi_first-principles_2003,ebert_anisotropic_2009}. In particular, the combinations of MFT with higher-level exchange-correlation approximations than the conventional LDA (local density approximation) or GGA (generalized gradient approximation) have been reported such as LDA+U, Hubbard-I and cluster DMFT (dynamical mean-field theory) \cite{solovyev_effective_1998-1,wan_calculation_2006}. It is important to have these capabilities because each approximation has its own validity limit to any of which a magnetic material can belong; {\it i.e.}, strongly or moderately correlated materials.

The main motivation of current study comes from the absence of the combination of MFT with GW method. GW is a well-established standard approximation going beyond LDA  and GGA. Although its simple description of electronic self-energy based on RPA (random phase approximation) is certainly limited for describing strongly correlated materials, GW method cannot only provide useful information for moderately correlated materials \cite{han_quasiparticle_2014,jang_quasiparticle_2015,ryee_quasiparticle_2016}, but it also be combined with other technique like DMFT (dynamical mean-field theory) \cite{georges_dynamical_1996,biermann_first-principles_2003} and EDMFT (extended DMFT) \cite{si_kosterlitz-thouless_1996,smith_spatial_2000,sun_extended_2002}. 
Further, we emphasize that the currently available GW formalism including its self-consistent version is not well suited for calculating total energy. This poses a serious problem when one tries to estimate the magnetic interaction within GW approximation because the total energy comparison is not feasible. While the implementation of MFT combined with GW method is formally straightforward just as its combination with DMFT \cite{wan_calculation_2006}, it has never been realized or reported to the best of our knowledge. 

In this paper, we report the successful implementation of MFT combined with quasiparticle self-consistent GW (QSGW) method \cite{vanschilfgaarde_quasiparticle_2006,kotani_quasiparticle_2007}. The benchmark calculations for
metallic magnetic systems as well as classical correlated insulators clearly show that the GW self-energy contributes to the magnetic interaction, and its effect can be quite significant in some cases.
We also explore the unique features of MFT such as the orbital resolution of magnetic interaction.
This combination of MFT with QSGW can serve as a useful tool to study various magnetic materials.


\begin{figure}
	\centering
	\includegraphics[width=0.8\linewidth]{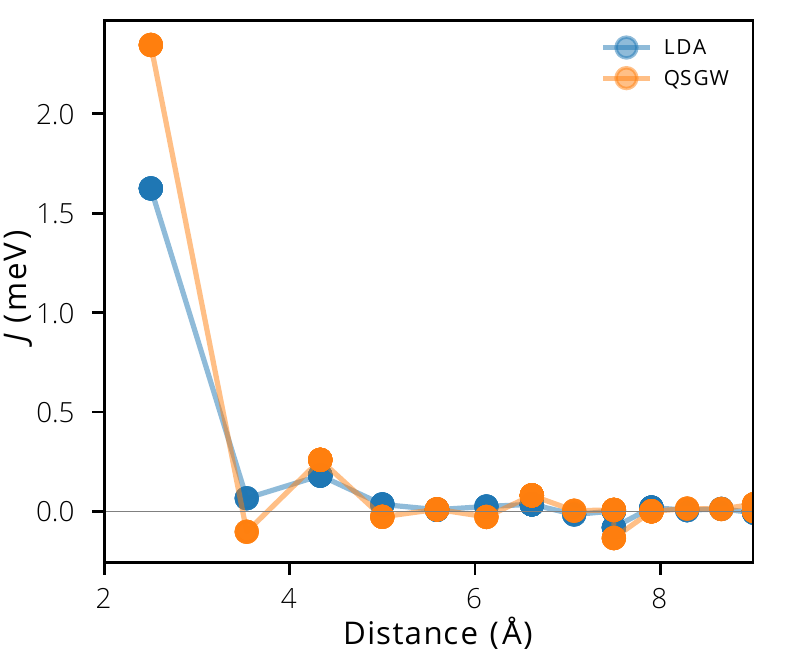}
	\caption{The calculated magnetic interaction for fcc Ni as a function of interatomic distance. The blue and orange circles represent the results of LDA and QSGW, respectively. 
		Note that the calculated spin moments are also different; LDA (0.64 $\mu_B$) and QSGW (0.78 $\mu_B$).}
	\label{fig:ni2j}
\end{figure}

\section{Computation Methods}

The QSGW method takes into account of the non-local self-energy $\Sigma \left( \mathbf { r } , \mathbf { r } ^ { \prime } , \omega \right)$ from dynamically screened Coulomb interactions within GW approximation. Starting from the non-interacting Hamiltonian $H_0$ ({\it e.g.}, from LDA), QSGW Hamiltonian $H_{\rm QSGW}$  is updated self-consistently by considering the screened Coulomb interactions $W$ within random phase approximation (RPA).
Hamiltonian including the $\omega$-dependent GW self-energy reads
$H(\omega) = \frac { - \nabla ^ { 2 } } { 2 m } + V ^ { \mathrm { ext } } + V ^ { \mathrm { Hart } } + \Sigma ( \omega )$ where $V ^ { \mathrm { ext } }$ and $V ^ { \mathrm { Hart } }$ is the external and Hartree potential, respectively. QSGW constructs the effective single-particle Hamiltonian by taking a static non-local exchange-correlation potential defined by $\begin{aligned} V ^ { \mathrm { xc } } & = \frac { 1 } { 2 } \int _ { - \infty } ^ { \infty } d \omega \mathrm { Re } [ \Sigma ( \omega ) ] \delta \left( \omega - H ^ { 0 } \right) + \mathrm { c.c. } \\ 
& =\sum_{ i j} \ket{ \psi_{i}} \bra{\psi_{i}} \frac { \mathrm { Re } \left[ \Sigma \left( \varepsilon _ { i } \right) + \Sigma \left( \varepsilon _ { j } \right) \right] } { 2 }   \ket{ \psi_{j}} \bra{\psi_{j}}.
\end{aligned}$\\
Here $\psi$ and $\varepsilon$ is the eigenfunction and eigenvalue, respectively, of the one-body effective Hamiltonian $H_{\rm QSGW}$. ${\mathrm {Re} } [ \Sigma ( \varepsilon _ { i } )]$ is the Hermitian part of the self-energy \cite{faleev_all-electron_2004,vanschilfgaarde_quasiparticle_2006}.
By self-consistently solving the Kohn-Sham equation with this effective potential, one can obtain
$H^{\rm QSGW}$, $\psi^{\rm QSGW}$ and $\varepsilon^{\rm QSGW}$.

Formally, combining MFT with QSGW is equivalent to that with DMFT which was reported by Wan \textit{et al.} \cite{wan_calculation_2006}; one can just replace DMFT self-energy by GW self-energy. Since the effect of GW self-energy is contained in the self-consistently determined $H^{\rm QSGW}$, $\psi^{\rm QSGW}$ and $\varepsilon^{\rm QSGW}$, the formal expression of MFT is unchanged.
The magnetic interaction between site $i$ and $j$ is therefore given by \cite{lloyd_multiple_1972,
		oguchi_band_1983,oguchi_magnetism_1983,liechtenstein_exchange_1984,liechtenstein_local_1987,udvardi_first-principles_2003,han_electronic_2004,mazurenko_weak_2005,ebert_anisotropic_2009,antropov_exchange_1997,yoon_reliability_2018}
	\begin{equation} \label{Eq_Jij_momentumspace}
	J_{ij}({\bf{q}} ) = \frac{1}{\pi} {\rm{Im}} \int_{}^{}  
	\int_{}^{\epsilon_{\rm{F}}}  d{\bf{k}} \, d\epsilon  
	\rm{\, Tr}[
	V_{{\bf{k}},i}^{\downarrow \uparrow } {\mathbf{G}}_{{\bf k},ij}^{\uparrow\uparrow}   
	V_{{\bf{k+q}},j}^{\uparrow \downarrow}  {\mathbf{G}}_{{\bf k+q},ji}^{\downarrow\downarrow} 
	]
	\end{equation}
	where the Green function is given in terms of eigenvector and eigenvalue. With orbital index $l$,
	\begin{equation} \label{Eq_green_DFT}
	\mathbf{G}^{\uparrow \uparrow {\rm (QSGW)}}_{l_1 l_2 ,{\bf{k}},ij } = \sum_{n}^{} \frac{ \ket{\psi_{l_1 ,{\bf{k}},i }^{\uparrow {\rm (QSGW)}}} \bra{\psi^{\uparrow {\rm (QSGW)}}_{l_2  ,{\bf{k}},j  } } }{z-\epsilon^{\uparrow {\rm (QSGW)}}_{n,{\bf{k}} }   + i\eta}.
	\end{equation} 
	For the case of collinear spins, the off-diagonal parts (${\bf{H}^{\downarrow\uparrow}}$ and ${\bf{H}^{\uparrow\downarrow}}$) are all zero. The perturbation term by the infinitesimal spin rotation is expressed by
	\begin{equation}\label{Eq_Vdef}
	V_{l_3 l_1,{\bf{k}},i }^{\downarrow  \uparrow {\rm (QSGW)}} = {\frac{1}{2}}( {\bf{H}}_{l_{3} l_{1},{\bf k},i }^{\downarrow  \downarrow {\rm (QSGW)}} - {\bf{H}}_{l_{3} l_{1},{\bf k},i }^{\uparrow \uparrow {\rm (QSGW)}} ).
	\end{equation}
	It is also straightforward to calculate the orbitally-decomposed magnetic interaction \cite{yoon_reliability_2018} within QSGW.
	Note that the expression of $J_{i,j}({\bf q})$ in Eq.~(\ref{Eq_Jij_momentumspace}) has four orbital indices; $l_1$ and $l_3$ belonging to the site $i$, and $l_2$ and $l_4$ to the site $j$. 
	The interaction between the two orbitals (say, $l_1$ and $l_2$) is then calculated by \cite{yoon_reliability_2018}:
	\begin{equation}\label{Eq_orbital_J}
	J_{ ij  }^{l_1,l_2} ( {\bf{q} } ) =\sum_{l_3,l_4}^{} J_{ij}^{l_1, l_2, l_3, l_4} ( {\bf{q} } ).
	\end{equation}
Throughout the manuscript, we used the following convention for spin Hamiltonian, 
\begin{equation} \label{Eq_Heisenberg_spin_Hamiltonian}
H=-\sum_{i\neq j} J_{ij} {\bf e}_i\cdot{\bf e}_j,
\end{equation}
where ${\bf e}_{i,j}$ refers to the unit spin vectors of atomic sites $i$ and $j$. Note that MFT procedure requires the eigenfunction information, and therefore, the one-shot GW cannot be utilized to perform MFT.

QSGW calculations have been performed with our `ecalj' software package \cite{noauthor_https//github.com/tkotani/ecalj_nodate,noauthor_http//www.lmsuite.org_nodate} which takes so-called `PMT' (augmented plane wave (APW) + muffin-tin orbital (MTO)) as a basis set \cite{kotani_fusion_2010,kotani_linearized_2013,kotani_quasiparticle_2014}. The number of ${\bf k}$ points for the first Brillouin zone is
12$\times$12$\times$12 and 8$\times$8$\times$8 for elemental transition metals and transition metal monoxides, respectively. The MTO radii used in our calculations are as follows: (i)
1.17, 1.21, and 1.21 {\AA} for elemental Fe, Co, and Ni, respectively. (ii) 1.22 and 0.93 {\AA} for Mn and O in MnO. (iii) 1.13 and  0.90 {\AA} for Ni and O in NiO. For comparison, we also performed LDA \cite{vosko_accurate_1980} and LSDA$+U$ calculations \cite{dudarev_electron-energy-loss_1998}. For all materials considered in this study, we used the
experimental lattice parameters \cite{furstenau_initial_1985,pearson_handbook_2013,antonov_electronic_2006,gong_ultrafine_1991,gong_ultrafine_1991}.

\begin{figure}[t] 
	\centering
	\includegraphics[width=0.99\linewidth]{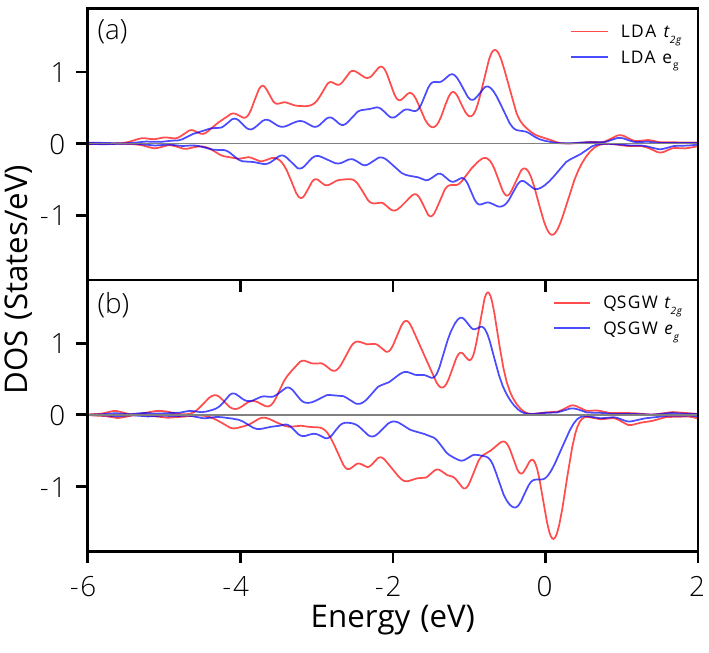}
	\caption{The calculated PDOS for the ferromagnetic Ni by (a) LDA and (b) QSGW. The red and blue lines represent the $t_{2g}$ and $e_{g}$ orbital states, receptively.}
	\label{fig:nitmdos}
\end{figure}

\section{Results and discussion}

Figure \ref{fig:ni2j} presents the calculated magnetic interaction strengths by QSGW for ferromagnetic fcc Ni as a function of inter-atomic distance. We emphasize that this kind of information can hardly be accessed because the total energy comparison is not feasible within the current first-principles GW schemes. By combining MFT with QSGW, the effect of GW self-energy on the magnetic interactions is now successfully estimated. Further, not depending on the model mapping, MFT calculates both short and long-range interactions from a single self-consistent calculation. Indeed, for this metallic magnet, the RKKY-like long-range feature is well observed.

It is instructive to compare the QSGW result (orange colored) with LDA (blue colored) in Fig.~\ref{fig:ni2j}. The overall interaction profile as a function of distance is quite similar in the two approximations. In fact, the calculated electronic structure by QSGW is not much different from that of LDA. Fig.~\ref{fig:nitmdos}(a) and (b) presents the projected density of states (PDOS) for Ni-$d$ states calculated within LDA and QSGW, respectively. A noticeable feature is the bandwidth reduction: In QSGW result, the overall bandwidth is slightly reduced and the prominent peaks are narrower for both $t_{2g}$ and $e_g$ states. It is the effect of GW self-energy and the similar features have also been observed in the other metallic systems  \cite{tomczak_many-body_2012,han_quasiparticle_2014,jang_quasiparticle_2015,ryee_quasiparticle_2016}. This change in the band structure is reflected in the results of magnetic interactions shown in Fig.~\ref{fig:ni2j}. The basically same feature is found also for other magnetic elements, Fe and Co (see Table~\ref{table_metal}).
It should be noted that the results presented in Fig.~\ref{fig:ni2j} is the $J$ values defined by Eq.~(\ref{Eq_Heisenberg_spin_Hamiltonian}).
The calculated moment is 0.64 $\mu_B$ in LDA and is enhanced 0.78 $\mu_B$ in QSGW being consistent with the previous report \cite{sponza_self-energies_2017}.
To conclude, in these metallic ferromagnets, the slight modification of band structure by GW self-energy causes a little change in the magnetic interactions.

\begin{table}[]
	\centering
	\begin{tabular}{l|c|c|cl} 
		\hline
		& Fe    & Co  &  Ni  &  \\ \hline
		& \multicolumn{1}{c|}{$J_{1}$, $J_{2}$, $J_{3}$}       & \multicolumn{1}{c|}{$J_{1}$, $J_{1'}$, $J_{2}$, $J_{3}$} &  \multicolumn{1}{c}{$J_{1}$, $J_{2}$, $J_{3}$} &  \\ \hline
		LDA      &  \multicolumn{1}{l|}{8.2, 3.2, 0.1} & \multicolumn{1}{l|}{5.7, 7.8, 0.7, 1.2}   &  1.7, 0.05, 0.1&  \\ 
		QSGW  &  \multicolumn{1}{l|}{7.7, 4.0, 0.3} & \multicolumn{1}{l|}{4.6, 8.5, 0.7, 2.2}   &  2.4, $-$0.1, 0.3 &  \\ \hline
	\end{tabular}\\[5mm]
	
	\begin{tabular}{l|cc|cc}
		\hline
		& Fe & &  Ni   &  \\  \hline
		&  $e_g$ & $t_{2g}$ &   $e_g$ & $t_{2g}$  \\  \hline
		LDA      &  & \multicolumn{1}{l|}{}    &  &  \\
		$e_g$    & 4.8 & \multicolumn{1}{r|}{4.3}  &   0.2 & 0.2  \\
		$t_{2g}$ & 4.3 & \multicolumn{1}{l|}{$-$4.2} &   0.2 & 1.0 \\ \hline
		QSGW      &  & \multicolumn{1}{l|}{}   &  &  \\
		$e_g$ & 5.5 & \multicolumn{1}{r|}{4.3}  &  0.2 & 0.3 \\
		$t_{2g}$ & 4.4 & \multicolumn{1}{l|}{$-$5.0} &  0.3 & 1.5 \\ \hline
		
	\end{tabular}
	
	\caption{\label{table_metal} (Upper panel) The calculated magnetic coupling constants $J$ for bcc Fe, hcp Co, and fcc Ni by LDA and QSGW (in the unit of meV). The first, second, and third nearest coupling is represented by $J_1$, $J_2$, $J_3$, respectively. For hcp Co, $J_1$ and $J_{1'}$ refers to the out-of-plane and the in-plane interaction, respectively.  (Lower panel) Orbitally decomposed first neighbor magnetic interaction ($J_1$). Since magnetic force responses are calculated for each orbital, in these transition-metal systems 5$\times$5 matrices are obtained. Here the results are presented in 2$\times$2 matrix forms for clarity. See Ref.~\citenum{yoon_reliability_2018} for further computation details.
	}
\end{table}

One useful feature of MFT is to provide the orbital-resolved information for the magnetic couplings \cite{yoon_reliability_2018,kvashnin_microscopic_2016}. This capability is also successfully combined with QSGW whose result is summarized in Table~\ref{table_metal}. Note that in this orbital-decomposed MFT, the magnetic coupling is expressed not by a single number but by a matrix. In case of Ni, for example, the dominant contribution comes from $t_{2g}$--$t_{2g}$ interaction. In Fe, one the other hand, not only $t_{2g}$--$t_{2g}$ but also $e_{g}$--$e_{g}$ and  $t_{2g}$--$e_{g}$ are all important. It is interesting to note that $t_{2g}-t_{2g}$ interaction is antiferromagnetic as first noted by Kvashnin {\it et al.} \cite{yoon_reliability_2018,kvashnin_microscopic_2016}.

\begin{table}[]
	\centering
	\begin{tabular}{lllll}
		\hline
		& MnO &   NiO &   &  \\ \hline
		LDA     &  $-$3.2    &   $-$4.3  &  &  \\
		LDA+U &  $-$1.5  &   $-$3.7  &   &  \\
		QSGW  & $-$1.0  &   $-$1.4  &  &  \\ \hline
		\\
	\end{tabular}
	
	\begin{tabular}{l|rrr|rrr}
		\hline
		& MnO  &  &  & NiO   &  &   \\ \hline
		& $4s$ & $t_{2g}$ & $e_{g}$ & $4s$ & $t_{2g}$ & $e_{g}$  \\ \hline
		LDA & & & & & &\\
		$4s$         & 0.0 & 0.0 & 0.5  & 0.0 & 0.0 & 0.0 \\
		$t_{2g}$ & 0.0 & $-$1.2 & 0.0  & 0.0  & 0.0  & 0.0 \\
		$e_{g}$  & 0.5 & 0.0 & $-$3.6  & 0.0 & 0.0 & $-$4.5 \\ \hline
		LDA+U & & & & & & \\
		$4s$        & 0.0   & 0.0     & 0.4  & 0.0 & 0.0 & 0.1 \\
		$t_{2g}$ & 0.0   &  $-$0.7  &   0.0 & 0.0  & 0.2 & 0.0 \\
		$e_{g}$  & 0.4   &  0.0  &   $-$2.2  & 0.1 & 0.0 & $-$4.1  \\ \hline
		QSGW  & & & & & & \\
		$4s$         & 0.0 & 0.0 & 0.3     & 0.0 & 0.0 & 0.0 \\
		$t_{2g}$ & 0.0 &  $-$0.6  &   0.0    & 0.0  &  0.0 &  0.0 \\ 
		$e_{g}$   & 0.3 &  0.0  &   $-$1.3 & 0.0 & 0.0  & $-$1.6 \\  \hline
	\end{tabular}
	\caption{\label{table_tmo} (Upper panel) 
		The calculated strongest (the second neighbor) interactions $J_2$ for MnO and NiO in the unit of meV.
		For LSDA+U calculation, we used $U$ parameter from  our constrained RPA calculation \cite{yoon_reliability_2018,jang_quasiparticle_2015}.
		(Lower panel) Orbitally decomposed second neighbor interaction $J_2$. Together with transition-metal $t_{2g}$ and $e_{g}$ orbitals, we also present the $4s$ component with 3$\times$3 matrices.} 
\end{table}

\begin{figure}
	\centering
	\includegraphics[width=0.99\linewidth]{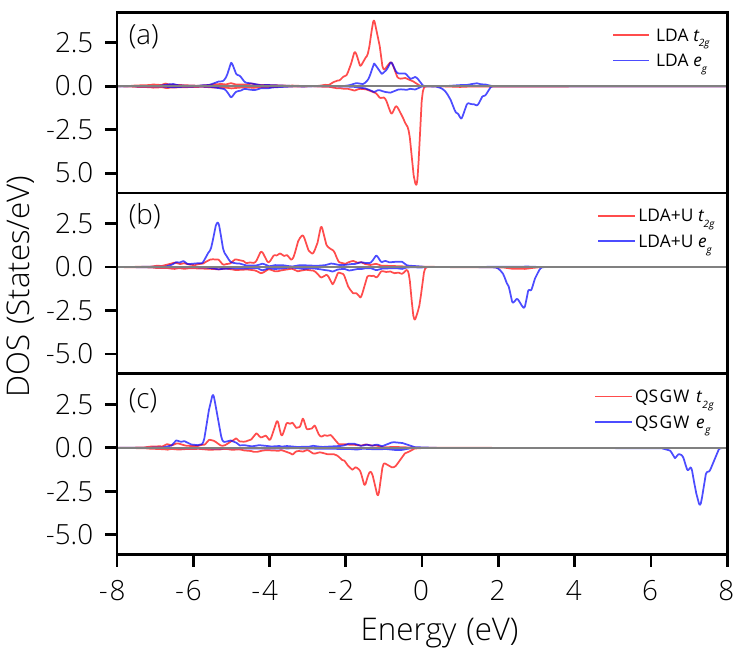}
	\caption{The Ni-$3d$ projected DOS for NiO calculated by (a) LDA, (b) LDA+U, and (c) QSGW. The red and blue lines represent $t_{2g}$ and $e_{g}$ states, receptively.
	}
	\label{fig:niotmdos}
\end{figure}

As the second examples, we considered a classical correlated insulators; namely NiO and MnO. The results are summarized in Table~\ref{table_tmo} where QSGW calculation is compared with LDA and LDA$+U$. The calculated magnetic coupling is significantly weaker in QSGW and LDA$+U$ than in LDA. It is attributed to the different electronic structure produced by three different approximations taken for electron correlations. The calculated PDOS for Ni-$3d$ are presented in Fig.~\ref{fig:niotmdos}; (a) LDA, (b) LDA$+U$ and (c) QSGW. The band gap is gradually and noticeably enlarged from LDA to LDA$+U$ and to QSGW while the basic electronic configuration, namely, fully-filled  $t_{2g}$ and half-filled $e_g$, is maintained in all cases. Since the superexchange for this second neighbor $J_2$ (the strongest interaction) is well approximated by $J_2\sim 1/\Delta$ ($\Delta$: band gap), the enlarged gap by introducing the additional self-energy naturally leads to the reduced magnetic coupling. This well-known feature has been observed in the previous LDA$+U$ and LDA+DMFT calculation \cite{oguchi_magnetism_1983,oguchi_transitionmetal_1984,han_electronic_2004,wan_calculation_2006,yoon_reliability_2018}. QSGW gives the even weaker magnetic interaction than LDA$+U$ for both MnO and NiO because it produces the larger band gap as clearly seen in Fig.\ref{fig:niotmdos}. It is known that QSGW can likely overestimate the bandgap \cite{aryasetiawan_electronic_1995,ku_band-gap_2002,kotani_quasiparticle_2007,sakuma_effective_2009,das_convergence_2015}
due to the RPA dielectric function neglecting the attractive interaction between electron and hole \cite{vanschilfgaarde_quasiparticle_2006,shishkin_accurate_2007}.

The orbital dependent magnetic interactions are presented in Table~\ref{table_tmo}.
As expected, $t_{2g}$-$t_{2g}$ interaction is sizable in MnO (high spin $d^5$ configuration) while $e_{g}$-$e_{g}$ interaction is dominant in NiO (high spin $d^8$). It is interesting to note that in MnO there is an interaction channel through Mn-$4s$ orbital. This interaction $J_{e_g-s}$ is sizable and ferromagnetic as shown in Table~\ref{table_tmo}. In the sense that the formally empty Mn-$4s$ orbital play a role in mediating magnetic interaction, this result is reminiscent of what was discussed in cuprate \cite{pavarini_band-structure_2001}.

\section{Summary}
By combining MFT with QSGW, we investigated the effect of GW self-energy on the magnetic interactions which can hardly be accessed due to the inability of current GW methodology to perform the total energy comparison of different magnetic phases. In metallic ferromagnetic $3d$ elements, GW self-energy slightly reduces the $d$ bandwidth and the magnetic interaction is enhanced accordingly while the same long-range feature is largely maintained. In the classical transition-metal monoxides, the magnetic interaction is significantly reduced by QSGW calculation due to the enlarged band gap. Some interesting features are found by the calculated orbital-resolved interaction profile which is a recently developed unique feature of MFT. For example, in MnO, the coupling between $e_g$ and the nominally-empty $4s$ orbital is noticeably large and comparable with $t_{2g}$--$t_{2g}$ coupling.


\section{Acknowledgments}
This work was supported by the Basic Science Research Program through the National Research Foundation of Korea (NRF)
funded by the Ministry of Education(2018R1A2B2005204) and Creative Materials Discovery Program through the NRF funded by Ministry of Science and
ICT (2018M3D1A1058754).


\bibliography{J_QSGW}

\end{document}